\documentclass[prl,twocolumn,superscriptaddress,showpacs]{revtex4}
\usepackage{graphicx}

\begin{document}

\title{Effects of Magnetic and non-Magnetic Impurities on the Superconducting State of YBa$_{2}$Cu$_{3}$O$_{7-\delta }$}
\author{M. Le Tacon}
\author{A. Sacuto}
\affiliation{Laboratoire Mat\'eriaux et ph\'enom$\grave{e}$nes Quantiques (UMR 7162 CNRS), Universit\'e Paris 7, 2 place Jussieu  
75251 Paris, France}
\affiliation{Laboratoire de Physique du Solide ESPCI, 10 rue Vauquelin 75231 Paris, France}
\author{Y. Gallais}
\affiliation{Departments of Physics and Applied Physics, Columbia University New York, NY 10027, USA}
\author{E. Ya. Sherman}
\affiliation{Department of Physics and Institute for Optical Sciences, University of Toronto, 
60 St. George St., Toronto M5S 1A7, Ontario, Canada}
\author{A. Forget}
\author{D. Colson}
\affiliation{Service de Physique de l'Etat Condens\'{e}e, CEA-Saclay, 91191 Gif-sur-Yvette, France}

\date{\today}

\begin{abstract}
We report electronic Raman scattering measurements on initially optimally doped Zn and Ni substituted Y-123 in A$_{1g}$ and B$_{1g}$ channels. We show that the B$_{1g}$ superconducting gap is independent of magnetic Ni and non-magnetic Zn impurity concentrations.
On the contrary, the energy of the A$_{1g}$ collective mode follows $T_c$ with 
two distinct slopes for Ni and Zn impurities, tracking the magnetic resonance detected by neutrons. 
We explain the unconventional behavior of the B$_{1g}$ superconducting
gap and discuss the evolution of the A$_{1g}$ mode under Ni and Zn impurities.
\end{abstract}
\pacs{74.62.Dh, 74.72.-h, 78.30.-j}
\maketitle
\noindent \qquad
%introduction
In the last few years, transport \cite{RA2003}, nuclear magnetic resonance
(NMR) \cite{Bobroff99}, muon spin relaxation ($\mu $-SR) \cite{Nachumi, Bernhard} 
and scanning tunneling microscopy (STM) \cite{Pan} measurements have shown that 
the effects of magnetic and non-magnetic impurities on the superconducting properties
 of cuprates are drastically different. Therefore they can be used as relevant probes for testing the quasiparticles and the collective modes in the superconducting state of cuprates.
Electronic Raman scattering (ERS) is a powerful tool for probing electronic excitations in selected areas of the Fermi surface. Raman responses are very sensitive to the $d$-wave character of the superconducting gap (SG). The B$_{1g}$ channel \cite{polar} probes the 
antinodal regions where the SG amplitude 2$\Delta$ is maximum while the B$_{2g}$ channel \cite{polar} probes the nodal regions where the SG vanishes. The A$_{1g}$ channel \cite{polar} has no symmetry restriction and is sensitive to nodal and antinodal regions of the Fermi surface. In the A$_{1g}$ channel an intense Raman active collective mode (the ''A$_{1g}$ mode''), which origin is not yet identified, is definitively observed in the superconducting 
state of optimally doped cuprates \cite{MoiMemeA1gB1g}, well below the 2$\Delta$ energy. 
It has been shown that the A$_{1g}$ mode tracks the magnetic 
resonance detected by inelastic neutron scattering \cite{Gallais2002} at 
\textbf{Q$_{AF}$} = ($\pi ,\pi $) for both its temperature and energy dependence 
under magnetic Ni substitutions.
% The A$_{1g}$ mode decreases following 5$k_{B}T_{c}$ under Ni
%substitutions.
Previous ERS studies on Zn substituted Y-123  with only one Zn concentration 
lead to contradictory results: on one hand the 2$\Delta$ energy of the SG seen in 
the B$_{1g}$ channel follows $T_c$ \cite{Limonov}, on the other hand it was argued to collapse to zero \cite{Martinho} when $T_c$ = 72 K.
Here we report ERS measurements on a wide concentration range of Zn impurities and compare them with those of Ni magnetic impurities. 
We show unambiguously that the 2$\Delta$ energy of the SG in B$_{1g}$ channel remains constant under magnetic and non-magnetic impurity substitutions up to 3\%, in contradiction with what we expect from a conventional superconductor. 
We show that the energy of the A$_{1g}$ mode follows $T_c$ with two distinct slopes for Zn and Ni impurities ans tracks the magnetic resonance energy does. Using an intermediate scattering potential with anisotropic phase shift for Ni impurities and an unitary limit scattering in the 
"Swiss Cheese"\cite{Nachumi} model for Zn impurities, we explain the unexpected  B$_{1g}$  SG energy dependence and discuss the A$_{1g}$ collective mode behavior.\par

Studied samples were optimally doped YBa$_{2}$(Cu$_{1-y_{\rm Zn}}$Zn$_{y_{\rm Zn}})_{3}$O$_{7-\delta }$ single 
crystals (grown by the ''self-flux'' method \cite{Kaiser87}) with various Zn concentrations:
$y_{\rm Zn}$=0 (\textit{T}$_{c}^{\rm (onset)}$=92.5K; $\Delta$\textit{T}$_{c}\sim2$ K), $y_{\rm Zn}$=0.3\% (87.5K; 2K), $y_{\rm Zn}$=0.7\% (83K; 3K), $y_{\rm Zn}$=1.5\% (73K; 4K) and $y_{\rm Zn}$=2\% (64K; 7K).
They have been renamed Y-123, Y-123:Zn87K, Y-123:Zn83K, Y-123:Zn73K and Y-123:Zn64K respectively.
Similarly, the Ni substituted Y-123 crystals of \cite{Gallais2002}, have been renamed
Y-123:Ni87K and Y-123:Ni78K.
Impurity concentrations were checked by chemical analysis using an electron probe.
\textit{T}$_{c}^{\rm (onset)}$ and 10\%-90\% width $\Delta$\textit{T}$_{c}$ have been evaluated
from ZFC DC-magnetization under a 10G field. The broadening of transition under impurity 
substitutions is consistent with previous works (see e.g. Ref. \cite{Fong} and references therein).
Zn and Ni are divalent ions known to offer a particularly attractive way to reduce $T_c$ without changing the carrier concentrations \cite{Mahajan94Bobroff97}.\par
ERS have been carried out with a T64000 JY spectrometer in triple subtractive configuration.
Crystals were mounted on the cold finger of an He circulation cryostat and cooled down to 10 K. 
The 514nm excitation line was used. The laser power on the crystal surface was kept below 
3mW to avoid significant heating (which is smaller than 3K according to Stokes-anti-Stokes ratio).
Raw spectra have been corrected for the Bose factor $n(\Omega,T)$ to get the imaginary part 
of the Raman susceptibility $\chi ^{\prime \prime }(\Omega)=I(\Omega)/[1+n(\Omega,T)]$. %\cite{Sacuto2000}.

%spectres expérimentaux, description

\begin{figure}[ptbh]
\begin{center}
\hspace{-5mm} \includegraphics[width=0.95\columnwidth]{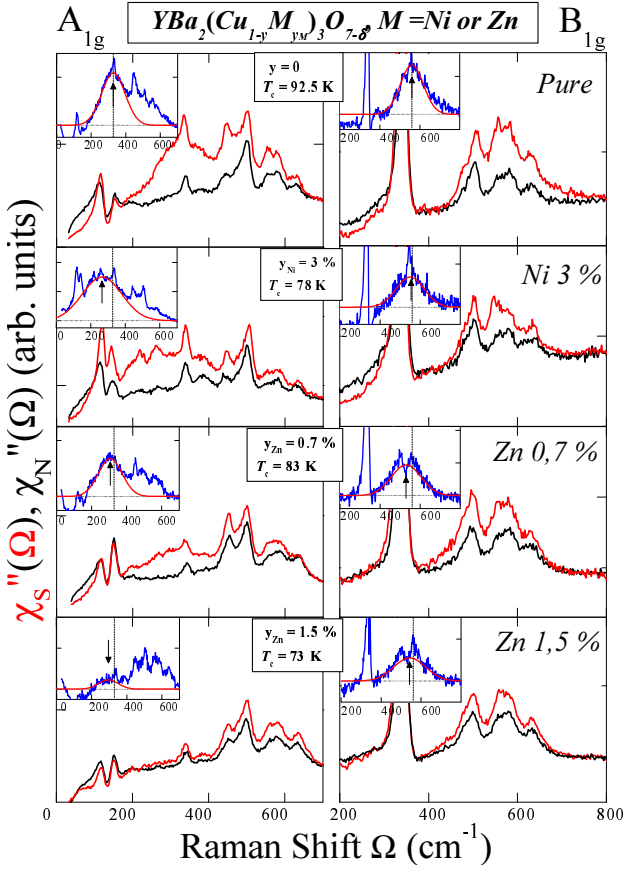}
\end{center}
%\par
\vspace{-4mm}
\caption{Superconducting and normal Raman responses $\chi^{\prime\prime}_{S}(\Omega)$ (in red)
and $\chi^{\prime\prime}_{N}(\Omega)$ (in black) of selected YBa$_{2}$(Cu$_{1-y_M}$M$_{y_M})_{3}$O$_{7-\delta }$ 
(M = Ni, Zn) single crystals. The difference $\chi^{\prime\prime}_{S}(\protect\Omega) - \protect\chi^{\prime\prime}_{N}(
\Omega)$ (in blue) is plotted in the insets.}
\label{fig:Zn}
\vspace{-4mm}
\end{figure}

Figure \ref{fig:Zn} shows the susceptibilities $\chi_N^{\prime \prime }(\Omega)$ 
(resp. $\chi_S^{\prime \prime }(\Omega)$) in normal (resp. superconducting) state in the A$_{1g}$ and B$_{1g}$ channels for selected concentrations (y$_{\rm Ni}$=3\%, 
y$_{\rm Zn}$=0.7\% and 1.5\%) of Ni and Zn impurities. The spectra related to y$_{\rm Ni}$=1\%, 
y$_{\rm Zn}$=0.3\% and 2\% are not shown here and can be found in Refs. \cite{MoiMemeA1gB1g, Gallais2002}.
The phonon lines for both A$_{1g}$ and B$_{1g}$ channel are well identified as well as additional weak features related to slight CuO chains disorder \cite{Gallais2002}.
For pure Y-123, we have a clear enhancement of the A$_{1g}$ and B$_{1g}$ electronic responses between (170-670 cm$^{-1}$) and (400-800 cm$^{-1}$) respectively in the superconducting state compared to the normal one. Subtractions of the normal contribution from the superconducting one are shown in insets. In B$_{1g}$ channel, we obtain a well defined peak centered at 565 cm$^{-1}$ (8.8 $k_{B}T_{c}$) corresponding to the 2$\Delta$ pair breaking energy of the SG. In the A$_{1g}$ channel, the subtraction gives rise to a broad asymmetric peak, in which two distinct contributions have already been established \cite{MoiMemeA1gB1g}. The first one is the A$_{1g}$ collective mode centered at 331 cm$^{-1}$
(5 $k_{B}T_{c}$), that runs from 170 to 410 cm$^{-1}$ and the second one, from 410 to 670 cm$^{-1}$ is the SG signature in the A$_{1g}$ channel \cite{MoiMemeA1gB1g}.

As Ni impurities are introduced the A$_{1g}$ mode softens and broadens but does not exhibit any significant decrease of its intensity up to y$_{\rm Ni}$=3\%. On the contrary the A$_{1g}$ mode under Zn impurities reduces to a weak contribution around 295 cm$^{-1}$ for y$_{\rm Zn}$ =1.5\% and totally disappears for y$_{\rm Zn}$=2\% \cite{MoiMemeA1gB1g}.
%For these two last concentrations, only the signature of the SG is clearly detected around 550 cm$^{-1}$. 
On the other hand, the B$_{1g}$ SG broadens but remains almost constant in energy with respect to Ni and Zn impurity insertions, in contrast with the A$_{1g}$ mode which shifts to lower energy.

\begin{figure}[tbh]
\begin{center}
%\vspace{-5mm}
\includegraphics [width=0.95\columnwidth, height = 6.5 cm] {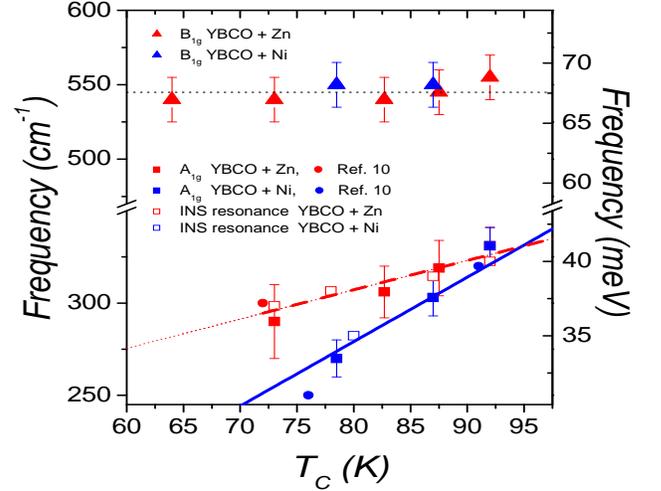}
\end{center}
\par
\vspace{-7mm}
\caption{
Energies of the SG peak (triangles), A$_{1g}$ mode (filled squares) and of the neutron resonance
(empty squares) with respect to \textit{T}$_c$ under magnetic Ni (blue) and non magnetic Zn (red) impurity substitutions. Data from reference \onlinecite{Martinho} have been added as filled circles. }
\label{fig:Bilan}
\vspace{-4mm}
\end{figure}

In Fig. \ref{fig:Bilan} are reported the A$_{1g}$ mode and the B$_{1g}$ SG energies with respect to $T_c$ for all Ni and Zn impurity concentrations. 
In addition to our measurements we have plotted the data of Ref.\cite{Martinho}.
In contrast to what we expect from a conventional superconductor, the $2\Delta$ SG energy remains almost constant as $T_c$ decreases with Ni and Zn impurities. 
In particular our experimental data show that the B$_{1g}$ SG energy in Y-123:Zn neither falls to zero as $T_c$ reaches 72 K (as claimed in Ref. \cite{Martinho}), nor scales with $T_c$ (see Ref. \cite{Limonov}) but remains constant in energy over a wide concentration range down to $T_c$ = 64 K. 
Its average value $\langle 2\Delta\rangle$ is $\sim 67.6$ meV ($\pm 2.5$ meV).
It has already been shown that changes of $T_c$ under oxygen doping fail to keep the $2\Delta/k_BT_c$ ratio constant, and our data show that even for fixed doping (here the optimal one), when $T_{c}$ decreases under magnetic or non-magnetic impurities substitutions, $2\Delta/k_{B}T_{c}$ is not constant anymore.

Figure \ref{fig:Bilan} shows that the A$_{1g}$ mode energy softens faster for Ni impurities than for the Zn one:
the energy of A$_{1g}$ mode follows 5$k_BT_c$ for Ni, whereas it follows, for Zn, $k_{B}(2.2T_{c}+2.8T_{c}^{\rm opt})$, 
where $T_c^{\rm opt}$= 92.5 K is the optimal $T_c$ of Y-123.
We notice that the data of Ref.[\onlinecite{Martinho}] follows the two distinct A$_{1g}$ slopes for Ni and Zn impurities.
This confirms that the ''A$_{1g}$-mode-energy/$T_c$'' ratio is not constant (and not equal to 5$k_B$) for Zn substitutions.

From STM measurements it has been shown that the quasiparticle scattering on Ni impurities
is predominantly potential, rather than magnetic \cite{Hudson}.
Based on this finding, we have calculated the superconducting density of states (SDOS) and the $B_{1g}$ Raman response for a CuO$_2$ plane where Cu is substituted by Ni. Considering potential pair breaking in a {\it d}-wave model, described by the usual Nambu-Gorkov' Green functions (see e.g., Ref.[\onlinecite{bouquinRusse}])
The SDOS and the imaginary part of the $B_{1g}$ Raman response are given by:
\begin{equation}
N_S(\Omega) = -\frac{1}{\pi}\Im \sum_{\mathbf{k}}{\rm Tr} \hat{G}({\mathbf k},i\omega_n)
{|_{i\omega_n\rightarrow \Omega+i0^{+}}},
\end{equation}
and
\begin{eqnarray}
\chi_{B_{1g}}^{\prime \prime }(\Omega) & = & -T\Im \sum_{{\mathbf k}}\sum_{\omega_n}
\gamma_{B_{1g}}(\mathbf{k})^2{\rm Tr}\big[\hat{\tau}_3\hat{G}(\mathbf{k},i\omega_n) \nonumber \\
	&	&\times \hat{\tau}_3\hat{G}(\mathbf{k},i\omega_n-i\Omega_{m})\big]{|_{i\Omega_{m}\rightarrow
	\Omega+i0^{+}}},
\end{eqnarray}
where $\gamma_{B_{1g}}(\mathbf{k)}$ is the $B_{1g}$ Raman vertex calculated in the effective mass approximation, and $\omega_n$ and $\Omega_m$ are fermionic and bosonic Matsubara frequencies, respectively. The details of calculation will be published elsewhere.
Scattering in the Born ($\delta$ = 0) and unitary ($\delta = \pi$/2) limits 
have already been treated in previous works \cite{Norman, Tom}. As scattering on Ni impurities is between these two limits \cite{Hudson}, we have opted for an intermediate momentum dependent phase shift $\delta(\mathbf{k})$, due to the anisotropy of the Fermi velocity $v_F$.
The self energy $\Sigma(i\omega_n)$ is then given by \cite{bouquinRusse}: 
\begin{equation}
\Sigma(i\omega_n) = \frac{n_{\rm imp}\sum_{\mathbf{k}} {\rm tan}(\delta(\mathbf{k}%
))^2{\rm Tr}\hat{G}(\mathbf{k},i\omega_n)}{(\pi N_F)^2- [\sum_{\mathbf{k}}{\rm tan}(\delta(%
\mathbf{k})){\rm Tr}\hat{G}(\mathbf{k},i\omega_n)]^2}
\end{equation}
where $n_{\rm imp}$ stands for Ni impurity concentration, and $N_F$ for
the DOS at the Fermi level in the normal state.
 
\begin{figure}[tbh]
\begin{center}
%\vspace{-5mm}
\hspace{0mm} \includegraphics [width=0.90\columnwidth, height = 6.5 cm]{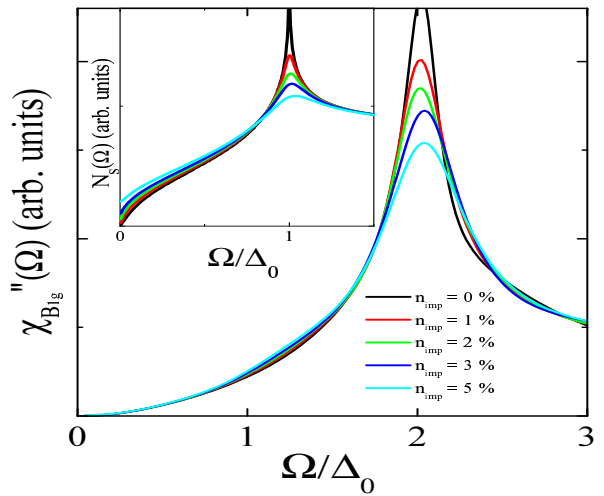} %, height = 5.8cm
\end{center}
\vspace{-7mm}
\caption{Evolution of the B$_{1g}$ Raman response as a function of impurity concentration for a {\it d}-wave superconductor.
$\Delta_0$ is the energy of the SG in the pristine case. The SDOS is plotted in inset.}
\label{fig:DOS}
\vspace{-6mm}
\end{figure}

In Fig. \ref{fig:DOS}, we have plotted the $B_{1g}$ Raman response as a function of Raman shift $\Omega$ for different $n_{\rm imp}$ (up to 5\%). The SDOS $N_S(\Omega)$ is plotted in inset.
At $\Omega=2\Delta_0$ (resp. $\Omega=\Delta_0$), with no impurities, $\chi_{B_{1g}}^{\prime \prime}$ (resp. $N_S(\Omega)$) diverges logarithmically. 
As Ni impurities are inserted, both $\chi_{B_{1g}}^{\prime \prime}$ and $N_S$ exhibit a maximum at the same energy than in the pristine case, suggesting that the quasiparticles in the antinodal regions are not affected by impurity scattering.
The intensity of $\chi_{B_{1g}}^{\prime \prime}(2\Delta_0)$ decreases with increasing $n_{\rm imp}$, just as observed experimentally. 
A finite DOS appears at the Fermi level $\Omega=0$, and increases with $n_{\rm imp}$. 
Low energy excitations in the superconducting state come from quasiparticles of the nodal regions. 
$N_S(\Omega=0)$ grows when $n_{\rm imp}$ increases implies that superconductivity is destroyed around the nodes of the SG, where the interaction leading to the Cooper pairs formation is weaker.
This is consistent with B$_{2g}$ Raman data \cite{Gallais2003} which reveals no distinctions
between superconducting and normal B$_{2g}$ responses as from 1\% of Ni impurities.
This explains also why the B$_{2g}$ response is more generally affected by any scattering source
(e.g. structural disorder) as previously observed \cite{Gallais2003, Gallais2005, Nem}.
The situation is different for non-magnetic impurities since the
scattering has been found to be almost unitary ($\delta \sim 0.48\pi$
\cite{Pan}). As a consequence, the superconducting order parameter is suppressed 
locally around each impurity site, and the superfluid density decreases. This gives rise to 
the "Swiss Cheese" model and explains why the SG intensity decreases as Zn 
concentration increases.
To achieve the calculation of DOS and Raman response in a Zn substituted cuprate, 
a serious difficulty arises from the fact that one must take into account spatial inhomogeneities of the superconducting order parameter. It has not been performed here. 
However, our data suggest that, in the remaining superconducting condensate, 
the Cooper pairs binding energy is not altered, and thus no changes are seen in the pair breaking peak in the B$_{1g}$ channel. 
 
At this step, it is interesting to notice that non-magnetic impurity effects in cuprates, at least on transport and $T_c$, can be reproduced by defects induced by electron irradiation \cite{RA2003}. Angle resolved photoemission spectroscopy (ARPES) measurements performed on such
irradiated samples (Bi-2212 with a \textit{T}$_{c}$ down to 62 K) have revealed no changes
in the SG energy with respect to \textit{T}$_{c}$  \cite{Vobornik}. This is consistent with our data. In Fig. \ref{fig:Bilan}, we have added to the A$_{1g}$ mode energies under Ni and Zn substitutions, the magnetic resonance energies obtained from \cite{Sidis2000_Zn}. %, Sidis2000}.
This shows unambiguously that for these two substitutions, the A$_{1g}$ mode tracks the magnetic resonance. As a consequence, magnetic and non-magnetic impurities lead to two different slopes for both the A$_{1g}$ mode and the magnetic resonance.\par

\begin{figure}[tbh]
\begin{center}
\hspace{-5mm} \includegraphics [width=0.95\columnwidth, height = 6.5 cm] {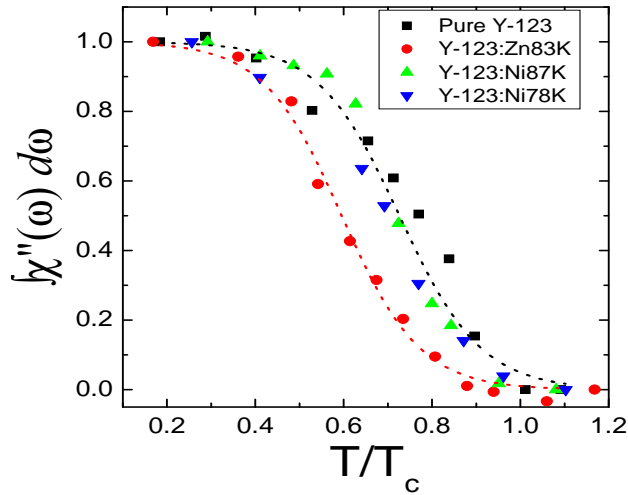} %
\end{center}
\par
\vspace{-6mm}
\caption{Temperature dependences of the spectral weight between 170 cm$^{-1}$ and 420 cm$^{-1}$of the A$_{1g}$ mode. The temperature scale has been normalized to $T_c$ for each crystal.}
\vspace{-5mm}
\label{fig:Tdep}
\vspace{0mm}
\end{figure}

To go further, we have plotted in Fig. \ref{fig:Tdep} the temperature dependences of 
the A$_{1g}$ mode for four crystals Y-123, Y-123:Zn83K, Y-123:Ni87K and Y-123:Ni78K.

We clearly see that the A$_{1g}$ mode starts growing just below $T_c$ in Y-123:Ni, in the
same way as pure Y-123. On the contrary, in the Zn case the A$_{1g}$ mode exhibits a relative
temperature delay of 0.8 $k_BT_c$ ($\sim$ 20 K). This has also been observed in Ref. [\onlinecite{Martinho}].
For Ni impurities, the A$_{1g}$ mode spectral weight follows the one of the free Y-123 as function of temperature.
The energies of both A$_{1g}$ mode and magnetic resonance follow 5 $k_BT_c$ for Ni-substituted and pure Y-123 
(Fig. \ref{fig:Bilan}), and STM measurements have shown the presence of localized states on the Ni impurity, but no disruption of superconductivity near the impurity \cite{Hudson}. 
These three experimental observations suggest that superconductivity is homogeneously modified 
by the introduction of magnetic impurities, and that the mechanism that leading to the formation of both magnetic resonance and A$_{1g}$ mode is not fundamentally affected, but rather
renormalized with respect to $T_c$. 
The scenario is different for Zn impurities.
On one hand, the shift in energy for both the A$_{1g}$ mode and magnetic resonance
is smaller than for Ni impurities (Fig. \ref{fig:Bilan}).
On the other hand, (see Fig. \ref{fig:Zn}) the intensity of the A$_{1g}$ 
mode decreases strongly with increasing Zn concentration, and for a concentration greater
 than 2\% this mode disappears, in contrast with Ni substitutions \cite{Gallais2002}. 
Finally, Fig. \ref{fig:Tdep} shows that the A$_{1g}$ mode starts growing 20 K below $T_c$ 
in the Y-123:Zn83K crystal.
Consequently, there exist Zn concentrations where superconductivity has settled, but the A$_{1g}$ mode is already not present.
This brings us to the conclusion that the diminution of superfluid density observed in 
$\mu$-SR \cite{Nachumi, Bernhard} for Zn-substituted samples does not explain alone the disappearance of the A$_{1g}$ mode.
As it has been shown previously, AF fluctuations at ${\bf Q}_{AF} = (\pi,\pi)$ are strongly enhanced above 
$T_c$ in Zn-substituted samples \cite{Sidis2000_Zn}, and survive in the superconducting state.
We believe that these fluctuations put back the A$_{1g}$ mode and the
neutron resonance \cite{Sidis2000_Zn}, and that the disappearance of these two collective modes when Zn concentration is larger than 2\% results from a conjugate effect of AF fluctuations and the diminution of the superfluid density.

In conclusion, we have shown that the unconventional behavior of the B$_{1g}$ SG
can be explained under anisotropic potential scattering for Ni magnetic substitutions.
Through Zn non-magnetic substitutions, we have confirmed the strong link between the 
magnetic resonance and the A$_{1g}$ mode. The A$_{1g}$ mode in Zn substituted Y-123 exhibits a
significant temperature delay from $T_c$ in comparison to the one in the Ni substituted Y-123 due to the presence of strong antiferromagnetic fluctuations.

\textbf{Acknowledgments:} We thank S. Paihl\`es, Y. Sidis, Ph. Bourges, M. Cazayous, 
K. Behnia, G. Deutscher, and P. Monod for very fruitful discussions.


\begin{references}
\bibitem{polar} B$_{2g}$ and B$_{1g}$ channels are obtained from cross polarizations of the incident and
scattered electric fields along and at 45$^{\rm o}$ from the Cu-O bounds
of CuO$_2$ layers. Parallel polarizations at 45$^{\rm o}$ from the Cu-O bounds give the A$_{1g}$+B$_{2g}$
channel, and pure A$_{1g}$ channel is obtained by subtracting the B$_{2g}$ one from the A$_{1g}$+B$_{2g}$ one.
\end{references}
\end{document}